# Detection of Auditory Brainstem Response Peaks using Image Processing Techniques in Infants with Normal Hearing Sensitivity

Amir Majidpour[1], Samer Kais Jameel[2], Jafar Majidpour[3], Houra Bagheri[1], Tarik A.Rashid[4], Ahmadreza Nazeri[1], Mahshid Moheb Aleaba[5]

[1] Audiology Department, School of Rehabilitation, Shahid Beheshti University of Medical Sciences, Tehran, Iran.
[2] Computer Center Department, Presidency University, Al-Karkh University of Science, Baghdad, Iraq.
[3] Department of Computer Science, University of Raparin, Rania, Iraq.
[3] Computer Science and Engineering Department, University of Kurdistan Hewlêr, Iraq.
[5] Department of Audiology, School of Rehabilitation, Arak University of Medical Sciences, Arak, Iran.

## Abstract

**Introduction**: The auditory brainstem response (ABR) is measured to find the brainstem-level peripheral auditory nerve system integrity in children having normal hearing. The Auditory Evoked Potential (AEP)s are generated using acoustic stimuli. Interpreting these waves requires competence to avoid misdiagnosing hearing problems. Automating ABR test labeling with computer vision may reduce human error.

**Method**: The ABR test results of 26 children aged 1 to 20 months with normal hearing in both ears were used. A new approach is suggested for automatically calculating the peaks of waves of different intensities (in decibels). The procedure entails acquiring wave images from an Audera device using the Color Thresholder method, segmenting each wave as a single wave image using the Image Region Analyzer application, converting all wave images into waves using Image Processing (IP) techniques, and finally calculating the latency of the peaks for each wave to be used by an audiologist for diagnosing the disease.

**Findings**: Image processing techniques were able to detect 1, 3, and 5 waves in the diagnosis field with accuracy (0.82), (0.98), and (0.98), respectively, and its precision for waves 1, 3, and 5, were respectively (0.32), (0.97) and (0.87). This evaluation also worked well in the thresholding part and 82.7 % correctly detected the ABR waves.

**Conclusion**: Our findings indicate that the audiology test battery suite can be made more accurate, quick, and error-free by using technology to automatically detect and label ABR waves.

## 1. Introduction

Auditory Evoked Potential (AEP) is an activity within the auditory system (ear, auditory nerve, and auditory areas of the brain) evoked by an acoustic stimulus. In other words, AEPs are brain waves (electrical potentials) generated using acoustic stimuli. Various stimuli including clicks, tones, and even speech sounds arouse these potentials (1, 2). Clinical applications of these responses include estimating hearing sensitivity in a particular population of adults and newborns at risk for hearing loss (1, 3), diagnosis of inner ear disease (e.g. Meniere's disease), tumor detection, and other impairments of the central auditory nervous system, and monitoring the central nervous system during nerve surgery (1, 4).

ABR test is a type of assessment of AEPs that is routinely used in the clinic to assess the integrity of the auditory pathway to the brainstem (5, 6). This test is a valuable tool in assessing hearing performance in clinical practices. A wide variety of stimuli elicit these responses from specific cochlear regions (7). Hayes and Jerger, in 1982 reported a relatively good consensus between ABR testing and behavioral thresholds with 2000 and 500 Hz tone-burst stimuli (8). This test is based on the voltage difference between the electrodes connected in different areas of the skull surface (9). 5 to 7 waves from the vortex (I-VII) can be recorded in the first ten milliseconds recording the test results. The recorded set of these waves forms the ABR results. Waves II, IV, and VI may not be detectable in people with normal hearing (10).

There is a great deal of variation in the ABR test results. The brainstem auditory response waves are quite different from other people and are similar to fingerprints in this respect. Moreover, it is infrequently possible to record the same ABR from two different people (1).

ABR test results are typically recorded and analyzed by a certified audiologist. In the neurological evaluation of the auditory system, this test is evaluated at the supra-threshold level. An average of the responses is taken to record a strong response with a good signal-to-noise ratio (11). This supra-threshold intensity is lowered in 10-20 dB steps until the V-wave is no longer detectable. The last intensity at which the V-wave was observed is considered as hearing threshold of the test subject. At the threshold level, the reproducibility is rechecked to ensure that the answer is correct. The general approach to analyzing ABR results is to manually determining the peak of the waves. The latency and amplitude of the peaks and the time interval between them are measured manually by an audiologist to estimate the amount of sound transmission to the brainstem. Therefore, the proposed method provides an automatic detection of ABR waves using Computer Vision techniques that helps to improve wave separation and thresholding.

Analyzing and interpreting ABR results requires a lot of knowledge and training (12, 13). Audiologists have shown a good level of agreement between inter-and intra-readers, and it has been found that experienced audiologists are more accurate in analyzing ABR test results. Those with less experience may, in some cases during tracking the threshold, make mistakes in detecting waves and subsequently give incorrect results (14). These fallacies may cause errors in the labeling of the waves and may cause errors in the final data analysis. Therefore, the automatic detection of ABR waves using Computer Vision techniques helps to improve wave separation and thresholding (15, 16). These new labeling results using computer methods will help audiologists make their final judgments about challenging patients (17, 18).

Few articles have used Computer Vision techniques in the field of hearing science and are only limited to a few outward features such as the diagnosis of the internal auditory canal and its nerves (19), the diagnosis of transient hearing loss in people with perforated tympanic membrane (20, 21), the use of IP techniques in spectrographs (22), facial temperature changes due to acoustic stimuli (23), congenital external auditory canal stenosis (24), automatic sign language detection for communication (25). Little research has been done in the field of ABR threshold tracking.

Thus, the main objective of this paper is to provide a tool that can be used in the audiologist test battery suite to improve accuracy, increase speed, and reduce human errors.

However, the contributions of the proposed method can be listed as:

1) IP techniques are used to accurately calculate the ABR waves.
2) The speed of interpretation of data is increased using computer-aided diagnosis schemes, this helps audiologists to improve their final clinical decisions.

The rest of the paper is organized as follows: Methodology is explained in section 2. Section 3, presents the most important results. In section 4, the discussion is added and finally, the key points in this paper are concluded in section 5.

## 2. Methodology

Information was gathered from newborns with normal hearing who underwent auditory electrophysiology in this study. The audiology group (conventional method) and a group that utilized IP (proposed method) to identify the waves were then given the results of the unlabeled ABR tests. The audiology experts identified the waves based on their academic knowledge and clinical expertise. IP methods were utilized by the computer team, which first transformed the waves into digital form. Each patient's waves were then categorized based on their intensity, latency, and ear type (right and left). Then, IP techniques, including the acquisition of the wave images, segmentation, converting all wave images into waves, and calculation of the peaks was applied to obtain wave peaks (See Figure 1). The output of the hearing and computer work was given to the statistical analyst to obtain accuracy and precision. Algorithm 1 has clearly explained all the steps of the proposed method.

### 2.1 Dataset

Data were obtained from 26 infants from 1 to 20 month-age (mean age was 4.34 months, SD=4.01) with normal ABR evaluation indication (Out of the 26 infants, 17 were female). The ABR was recorded as part of the neurological test battery, using an auditory evoked potential equipment from the Granson-Stadler (GSI) Audera brand. A 100 μs alternative click stimulus was presented at 80 dB nHL at a rate of 27.71 clicks/sec. Stimuli were presented monaurally to the right and left ear via ER-3A (Etymotic Research) insert earphones. The ABR was recorded by placing three surface electrodes at midline forehead (Fz) and nontest ear mastoid (ground) positions and reference to the test ear mastoid. Click-evoked responses to 500-to-2000 repetitions were averaged for each response with two replications in the threshold level. The time window is 15 milliseconds, and depending on their age, a 30 to 1500 Hz filter has been applied. Two experienced audiologists marked the ABR peaks on the resulting waveform. In the conventional method, two experienced audiologists marked the ABR peaks on the resulting waveform. The labels were classified as neurologically normal or abnormal, and the V-wave was tracked to the threshold. Characteristics of the waveforms labeled as abnormal included delays in activity peaks, low amplitudes, poor morphology and/or poor replicability, or deterioration in responses with increases in acoustic stimulation rate.

## 2.2 Feature extraction, Data processing, and identification challenges

As illustrated in Figure 1, the proposed method for automatically labeling the peak of the waves captured from the patient consists of the following steps.

1. Acquiring wave images from an Audera device using the Color Thresholder method.
2. Segmenting each wave as a single wave image using the Image Region Analyzer application.
3. Converting all wave images into waves using Image Processing (IP) techniques.
4. Calculating the latency of the peaks for each wave to be used by an audiologist for diagnosing the disease.

Thus, the new approach calculates the peaks of waves of different intensities (in decibels) automatically, which improves accuracy, increases speed, and reduces human errors.

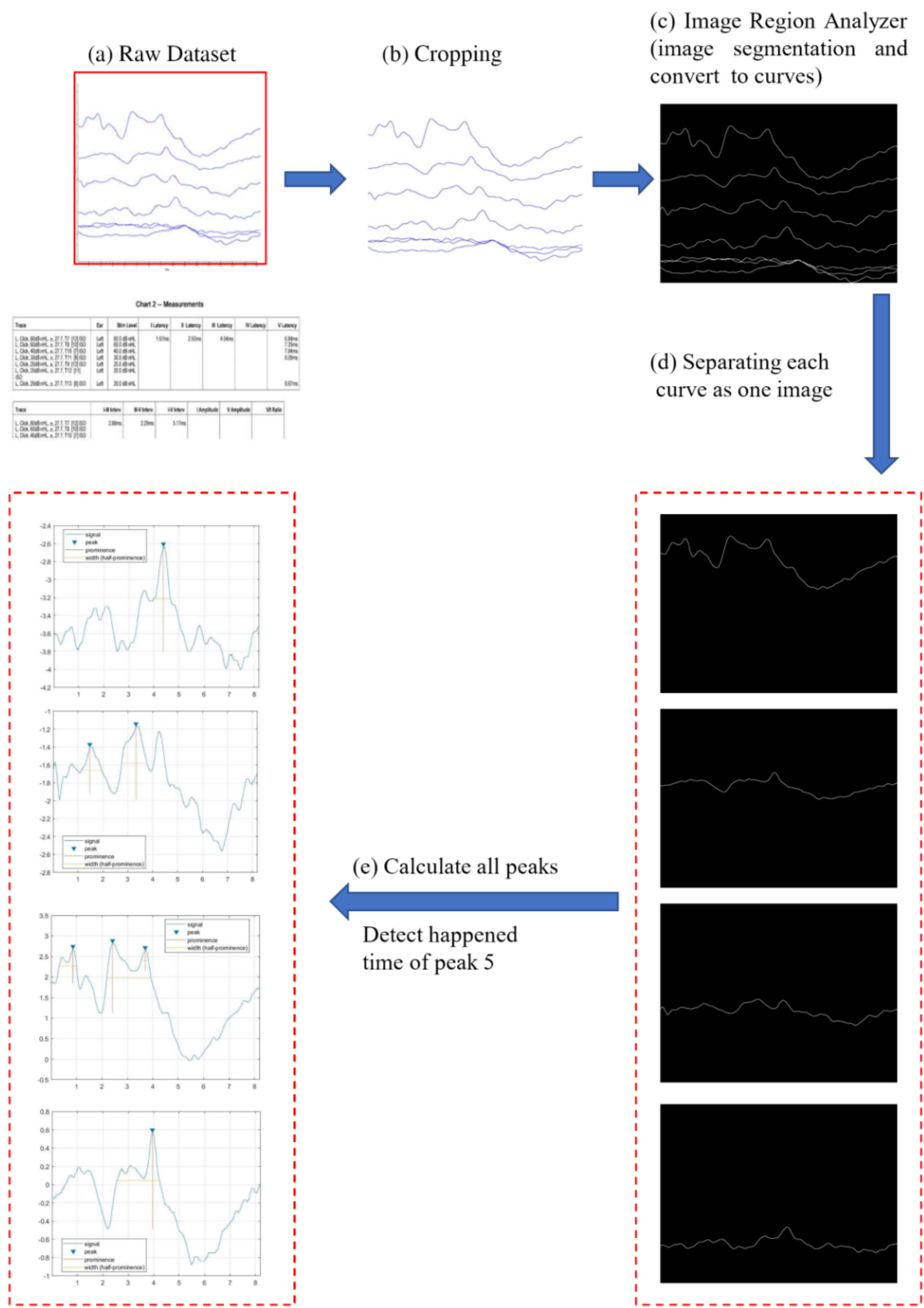

**Figure 1.** Stages of processing ABR data: (a) raw dataset; (b) cropping and resizing; (c) image segmentation based on the IRA method and conversion to curve; (d) separating each curve as one image; (e) calculating all peaks for each image.

**Algorithm 1.** Pseudo code of Proposed Method

| | |
|---|---|
| 1 | **Input**: *D: a raw dataset, img: a patient's image which is selected from the **D** (left or right ear);* |
| 2 | **For img=1:D** |
| 3 |    *img= Crop **img** to size (2351*1951)* |
| 4 |    *seg= Segment **img** to binary* |
| 5 |    *w=Recall Image Region Analyzer function for extracting each wave based on Minor Axis Length (**seg** as input)* |
| 7 |    **For i=1: w** |
| 8 |       *peak_wave= Calculate each peak of **i*** |
| 9 |       *time_wave= Calculate time of peak 5 for **i*** |
| 10 |       *return peak_wave, time_wave* |
| 11 |    **End for** |
| 12 | **End for** |
| 13 | **End** |

### 2.2.1 Acquisition of the wave images

An Audera device was used to record the ear patient's reaction as waves in various decibel (dB) levels on a sheet. To distinguish each side of a patient's ear, the device uses distinct wave colors: red for the right ear and blue for the left. Using the Color Thresholder technique, unique color is extracted for each side of the ear based on its color in the acquisition step. The Color Thresholder application divides color images into segments by thresholding the color channels using multiple color spaces; in this study, we employed the RGB color space, which means R refers to red, G refers to green, and B refers to blue color. It generates a color image's binary segmentation mask. The image, the three-color channels, and the color value of all pixels are shown as points in a 3-D color space plot by the application. By windowing the color channel values or creating a Region of Interest (ROI) in the image, it chooses the colors included in the mask (see Figure 2).

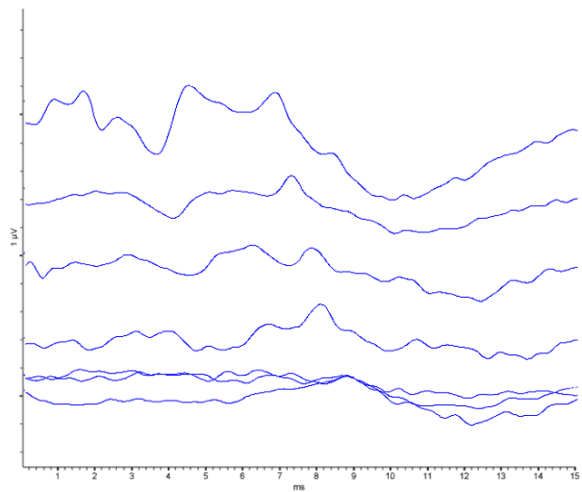

**Figure 2.** Remove unwanted information from ABR by crop it.

### 2.2.2 Segmentation

Image segmentation is a method of breaking down a digital image into several subgroups called Image segments to reduce the image's complexity and make further processing or analysis easier. In simple terms, segmentation is the process of assigning labels to pixels (26).

In this study, each wave image has a maximum of five waves with varied dB levels. In the segmentation process, we peruse at each wave separately to see how it may be segmented. The waves image may be used as an input, rather than analyzing the entire image, segmentation can be used to choose a section. This prevents the algorithm from analyzing the entire image, resulting in a faster inference time. As a result, all image elements (pixels) belonging to the same wave-type are given the same name. The Image Region Analyzer (IRA) application was used to perform the segmentation. The IRA application calculates and shows a set of attributes for each linked area (wave) as a table, as shown in Figure 3. The binary image and a table of attributes for each region are shown in the IRA app. Area, perimeter, orientation, major axis length, minor axis length, and other attributes of the image regions that were extracted and identified in the table's rows and columns, respectively, are shown in the table on the right side of Figure 3.

The Minor Axis Length (MLA) of the elliptical minor axis with the same region's normalized second central moments. The MLA parameter is utilized to build additional binary images by filtering the image on region properties, resulting in five wave images. (See figure 3).

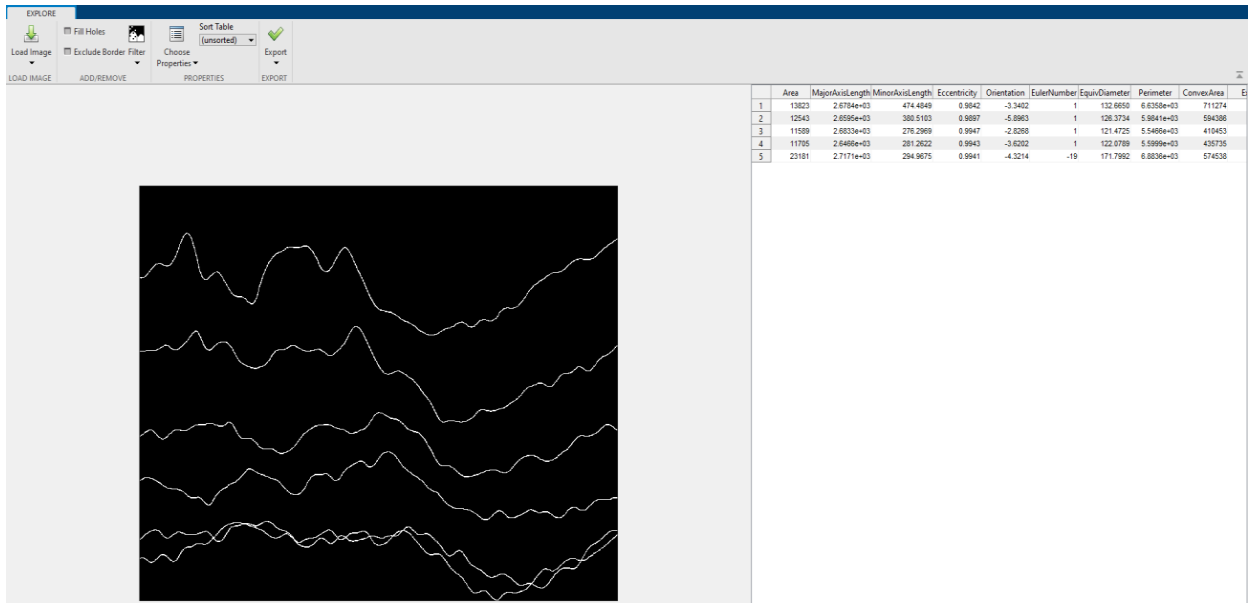

**Figure 3.** Separating all waves individually that can be process independently.

### 2.2.3 Convert all waves images into independent wave

The output from this stage makes it feasible to treat each wave independently without impacting the other waves, making this conversion stage one of the project's main stages. In this step, the wave image is converted into a wave that can be addressed as such by using the IP capability to capture data from image pixels and plot it on a 2-D line using the Plot function. In Figure 4, a segmented sample of a single wave from the original wave image is shown.

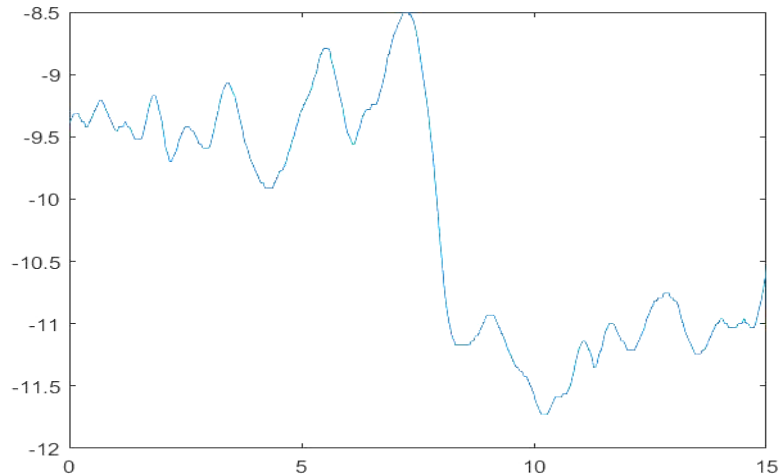

**Figure 4**. An output example of one wave processed independently from the rest of the waves (ABR).

### 2.2.4 Calculate the peaks

The find peaks function is used to locate and plot significant peaks; it returns a vector containing the input signal vector's local maxima (peaks). A local peak is a data sample that differs from the two samples adjacent to it in size. The peaks are output in the order of occurrence, as seen in figure 5. The function only returns the lowest index point if a peak is flat. The proposed method's structure is depicted in Figure 1. The procedures of proposed method are mentioned in algorithm 1.

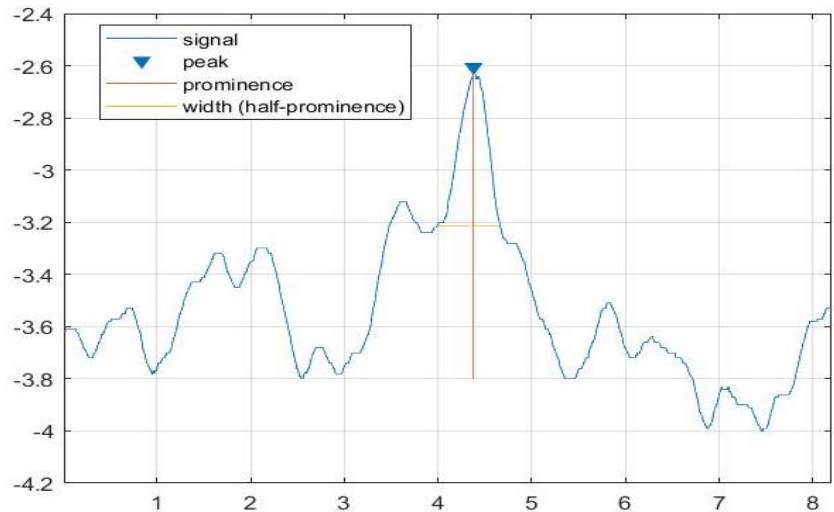

**Figure 5.** Calculate and located peaks in a wave.

### 2.3 Statistical analysis

First, the participants' demographic data were entered in Excel 2016 software. Then, for waves 1, 3, and 5 at the intensity level of 80 dB nHL and the fifth wave at intensities of 80, 60, 40, 30, and 20, Concordance Correlation Coefficient (CCC) was used to check the compatibility of the new method using IP techniques with the conventional method. According to the clinical application of the results of the evaluations, Bland–Altman plots were used to evaluate the discrepancies between the two methods, considering the conventional method as the gold standard. Finally, the correct and incorrect diagnoses and non-detection of the fifth wave at intensities of 80, 60, 40, 30, and 20 dB were calculated. Among the rationale for electing these particular waves, it can be posited that these three waves are the most widely encountered waves that lend themselves to the application of diagnostic parameters. Furthermore, waves III and V exhibit the greatest magnitude, thereby facilitating the identification of hearing thresholds at lower intensities (27). All calculations and output charts are performed using MedClac v20 software under Windows.

## 3. Result

Following data gathering, preliminary analyses were conducted to extract digital visual elements from the test results. These criteria included various types of hearing impairments and ABR wave properties such as latency and amplitude. The information from these samples was loaded into MedClac software to better understand the relationship between computer outputs and audiologist labels. We were able to get the anticipated modes' accuracy and precision.

**Table 1.** Accuracy and precision for image processing method based on clinical method in high intensity (80 dB nHL).

| Wave Type | 1 | 3 | 5 |
|---|---|---|---|
| Stimulus Intensity | 80 | 80 | 80 |
| Concordance correlation coefficient | 0.2688 | 0.9522 | 0.8564 |
| 95% Confidence interval | 0.02426 to 0.4830 | 0.9167 to 0.9728 | 0.7587 to 0.9165 |
| Pearson ρ (precision) | 0.3246 | 0.9706 | 0.8726 |
| Bias correction factor $C_b$ (Accuracy) | 0.8281 | 0.9811 | 0.9815 |

In Table 1, the CCC was used to check the compatibility of the two methods. The CCC results are presented for the three waves examined. The results indicate acceptable precision and accuracy for wave detection.

In this part of the data analysis, the accuracy of waves 5 (0.98) and 3 (0.98) at 80 dB is significantly higher, while wave 1 (0.82) has lower accuracy than other waves. Similarly, the precision of wave 1 (0.32) at 80 dB is low. However, the CCC for waves 3 (0.95) and 5 (0.87) is moderate to significantly high, showing the high ability of IP techniques in detecting waves. In addition to the results of the CCC to compare the new method with the conventional method, it should not be forgotten that the interpretation of the results of IP techniques requires a clinical overview, so Bland–Altman plots were used.

According to Graph 1, it can be seen that in detecting the location of the first wave at an intensity of 80 dB nHL, the average difference between the IP method and the conventional method is 0.05 seconds. The IP technique is a lower estimate for wave locations 1, 3, and 5 at 80 dB nHL, consistent with the CCC results. According to the confidence interval, the difference between the IP method and the conventional method with 95% confidence and in the intensity of 80 dB nHL stimulus for wave 1 is -0.39- + 0.3-millisecond, wave 3 is -0.25- + 0.11-millisecond, and for the wave 5 is -0.53- + 0.46.

The scatter of points in Graph 1 d`oes not show a specific difference in methods' results; it means that the IP techniques have a similar function in identifying the location of the waves at different period as conventional method.

**Graph 1.** Comparison of the performance of Bland–Altman latency between the conventional method and the IP techniques.

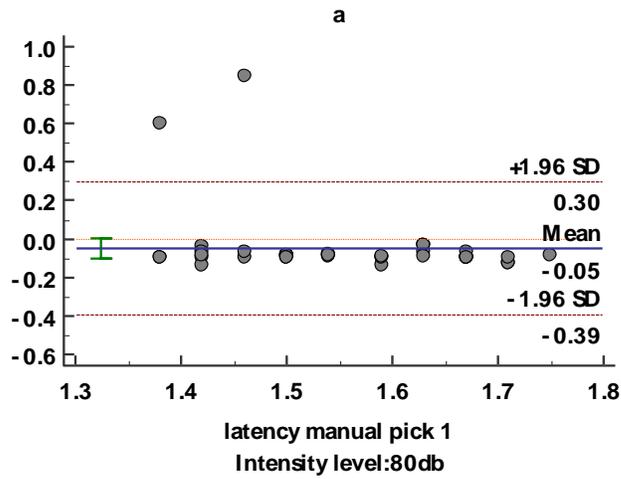

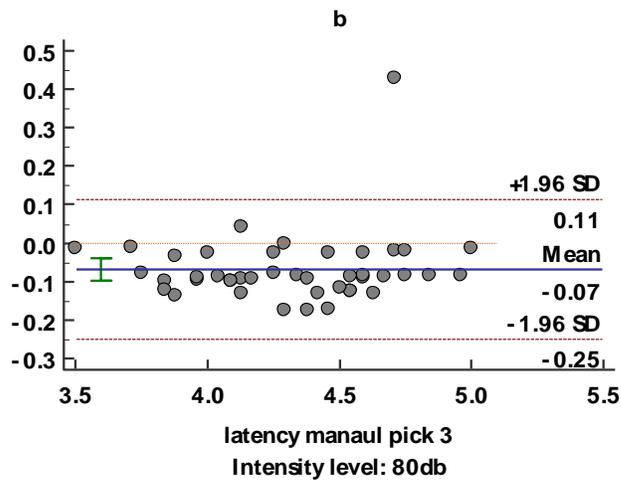

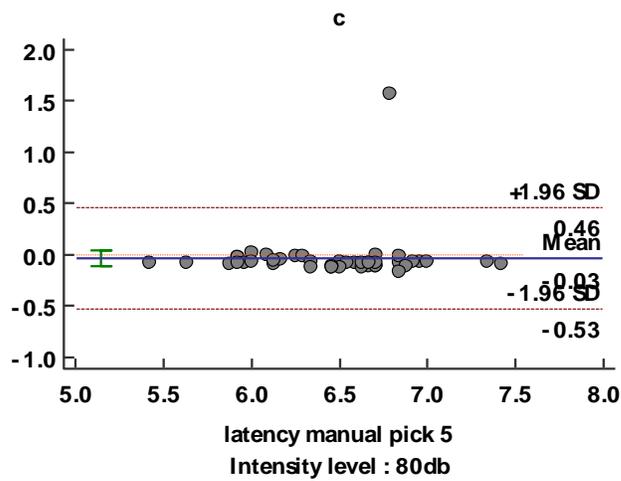

**Table 2.** Accuracy and precision for IP method based on clinical method in different intensities.

| Wave Type | 5 | 5 | 5 | 5 | 5 |
|---|---|---|---|---|---|
| Stimulus Intensity | **80** | **60** | **40** | **30** | **20** |
| Sample size | 43 | 42 | 50 | 50 | 49 |
| Concordance correlation coefficient | 0.8564 | 0.8907 | 0.3241 | 0.745 | 0.2139 |
| 95% Confidence interval | 0.7587 to 0.9165 | 0.8074 to 0.9392 | 0.09404 to 0.5213 | 0.6033 to 0.8411 | 0.02047 to 0.3920 |
| Pearson ρ (precision) | 0.8726 | 0.8935 | 0.3778 | 0.7831 | 0.3117 |
| Bias correction factor $C_b$ (Accuracy) | 0.9815 | 0.9969 | 0.8579 | 0.9514 | 0.6865 |

In Table 2, the CCC for wave 5 is calculated at varying intensity levels (80 dB to 20 dB). The detection of the fifth wave at intensities of 80 and 60 dB nHL has a high accuracy of 0.98 and 0.99, respectively. In contrast, the detection of the fifth wave at an intensity of 20 dB nHL has a low accuracy of 0.68, which needs further investigation. The point to consider is the existence of different precision that is observed in these analyzes. The highest precision is related to the intensity of 60 dB nHl (0.89), and the lowest value is related to the intensity of 20 dB nHl (0.31). At 40 dB, the accuracy of wave 5 detection is significantly lower (0.37), which requires more specialized investigation.

**Graph 2.** The percentage of accuracy of wave images detected at different intensities using IP.

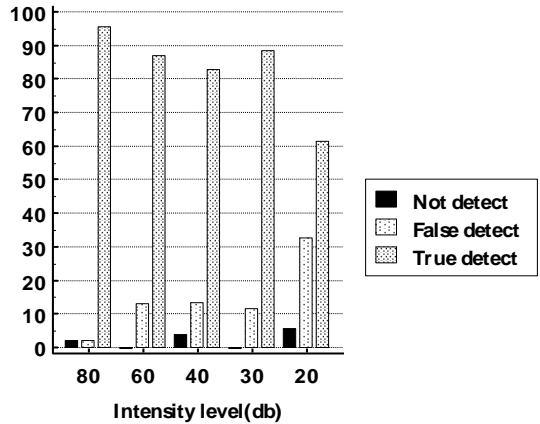

Graph 2 demonstrates that wave detection decreases with moving from high to low intensity. In general, the performance of this method in wave 5 detection was significantly high. This new method detected wave 5 with 82.7% (205) correctly in all intensities. In 14.9% (37), the IP misdiagnosed the waves and could not detect the wave in 2.4% (6). With a close look at the analysis, at 80 dB nHL, 95.65% of wave 5 were correctly detected. However, in contrast, at 20 dBs, it had a poor performance, so that only 61.53% of cases had a correct diagnosis, and in 32.69% of cases, it detected another wave instead of wave 5. In addition, in 5.76% of cases, it could not find wave 5. It means that the lower the intensities, the more difficult it is to detect waves for this tool, and of course, the error rate increases.

## 4. Discussion

By interpreting the findings of IP techniques, it can be concluded that this method can help detect the types of ABR waves that indicate neural integrity. Compared to the conventional peak selection method, IP techniques are the first study to use IP techniques to detect and track ABR waves obtained from infants.

The conventional method commonly used to detect and threshold ABR waves is based on compliance with predefined patterns (28). In pattern matching, the waves obtained are compared with the pattern generated by the average data of a normal population. Nevertheless, such averaging taken from the population may not be valid because the averaging action alters and delays peaks, amplitudes, and waveforms, leading to misinterpretation. Researchers can be informed that there are drawbacks to using this technique and suggest using more objective approaches.

Unlike threshold tracking, which uses multiple stimulation plates to assess a patient's hearing ability, the diagnostic field can use only a high-intensity level to assess the integrity of the nervous system (1, 11, 29). In the present study, an 80 dB intensity level was used to obtain a clear response with a good SNR for neurological detection.

The present study performed a large number of experiments with statistical assertions to create a robust and reliable tool that uses IP techniques. Based on the number of cases used in this assessment for the threshold section, the IP technique correctly detected 82.7% (205 waves), and only 2.4% (6 waves) could not detect wave 5, which can be attributed to noisy charts. 14.9% (37 waves) mistakenly identified another wave because there was no clear separation morphology between peaks 4 and 5 (30). In this study, it is clear that the amount of data studied is small, which significantly affects the performance of this tool. However, despite this limited amount of data, due to the high detection accuracy and the acceptable percentage of thresholding, this technique can be used as a valuable tool in the clinic to speed up the interpretation of data and enter the hearing test batteries.

Due to the clinical function of this tool and the possibility of improving wave detection, other researchers in this field are recommended to do more research to improve the detection and threshold of waves and merge IP techniques into the medical field.

## 5. Conclusion

The conventional method for detecting, labeling, and interpreting the ABR test results daily may not be handy for less experienced audiologists or even experienced audiologists with limited time. Even though little study has been done on hearing and computer vision, the IP approach's usefulness and practical relevance in improving the accuracy and speed of hearing examinations may be readily recognized. The current study employed a variety of IP approaches to identify ABR test waves better and determine the optimal techniques for improving wave labeling accuracy. As a result of these findings, a novel wave identification approach may be suited for developing an automated tool to detect ABR waveforms in the future. One of the weaknesses of the proposed method is the difficulty of segmenting the waves if some of them are intertwined. on the other hand, the input image must be clear and without impurities, otherwise, it will face the consequences of discovering the peak. In future work, in order to detect the wave peaks, we advise training a CNN model, which requires a large amount of data for it to perform well. On the other side, in the future, we are looking for comparisons with non-normal data in the study.